\documentstyle[iopconf1,epsfig]{article}
\begin{document}

\vspace{-1.cm}
\title{
Search for high energy neutrinos with the BAIKAL
underwater detector NT-96
}

\vspace{-1.cm}
\author{
V.A.Balkanov$^a$,I.A.Belolaptikov$^g$, L.B.Bezrukov$^a$, N.M.Budnev$^b$, A.G.Chensky$^b$,
I.A.Danilchenko$^a$, Zh.-A.M.Dzhilkibaev$^a$\footnote{E-mail:djilkib@pcbai10.inr.ruhep.ru}, G.V.Domogatsky$^a$, A.A.Doroshenko$^a$,
S.V.Fialkovsky$^d$, O.N.Gaponenko$^a$, O.A.Gress$^b$, D.Kiss$^i$, A.M.Klabukov$^a$,
A.I.Klimov$^f$, S.I.Klimushin$^a$, A.P.Koshechkin$^a$, Vy.E.Kuznetzov$^a$, V.F.Kulepov$^d$,
L.A.Kuzmichev$^c$, J.J.Laudinskaite$^b$, S.V.Lovzov$^b$, B.K.Lubsandorzhiev$^a$,
M.B.Milenin$^d$, R.R.Mirgazov$^b$, N.I.Moseiko$^c$, V.A.Netikov$^a$, E.A.Osipova$^c$,
A.I.Panfilov$^a$, L.K.Pan'kov$^b$, Yu.V.Parfenov$^b$, A.A.Pavlov$^b$, E.N.Pliskovsky$^a$,
P.G.Pokhil$^a$, E.G.Popova$^c$, V.V.Prosin$^c$, M.I.Rozanov$^e$, V.Yu.Rubzov$^b$, Yu.A.Seminei$^b$,
I.A.Sokalski$^a$, Ch.Spiering$^h$, O.Streicher$^h$, B.A.Tarashansky$^b$, G.Toht$^i$, T.Thon$^h$,
R.Vasiljev$^a$, R.Wischnewski$^h$, I.V.Yashin$^c$.}

\affil{$^a$\ Institute for Nuclear Research, 60-th October
Anniversary prospect 7a, Moscow 117312, Russia}

\affil{$^b$\ Irkutsk State University, Irkutsk, Russia}

\affil{$^c$\ Institute of Nuclear Physics, MSU, Moscow, Russia }

\affil{$^d$\ Nizhni  Novgorod  State  Technical University, Nizhni  Novgorod, Russia}

\affil{$^e$\ St.Petersburg State  Marine Technical  University, St.Petersburg, Russia}

\affil{$^f$\  Kurchatov Institute, Moscow, Russia }

\affil{$^g$\ Joint Institute for Nuclear Research, Dubna, Russia }

\affil{$^h$\ DESY-Zeuthen, Zeuthen, Germany} 

\affil{$^i$\ KFKI, Budapest, Hungary }

\vspace{-0.5cm}
\beginabstract
We present the results of a search for high energy neutrinos
with the Baikal underwater Cherenkov detector {\it NT-96.}
An upper limit to the flux of
$\nu_e+\nu_{\mu}+\bar{\nu_{\mu}}$ of
 $E^2 \Phi_{\nu}(E)<1.4\cdot 10^{-5}\,
\mbox{cm}^{-2}\,\mbox{s}^{-1}\,
\mbox{sr}^{-1}\,\mbox{GeV}$ is obtained, assuming an
 $E^{-2}$ behavior of the neutrino spectrum.
\endabstract

\section{Introduction}
The ultimate goal of large underwater neutrino telescopes
is the identification of extraterrestrial neutrinos of
high energy. In this paper we present results of a search for
neutrinos with $E_{\nu}>10 \,$TeV obtained with the deep underwater
neutrino telescope {\it NT-96} at Lake Baikal. 

The used search strategy for high energy neutrinos relies
on the detection of the 
Cherenkov 
light emitted by the electro-magnetic and (or) hadronic
particle cascades and high energy muons
produced at the neutrino interaction
vertex in a large volume around the neutrino telescope.
Earlier, a similar strategy has been used by the \mbox{DUMAND 
\cite{DUMAND}} and the \mbox{AMANDA \cite{AMANDA}}
collaborations to obtain upper limits on the diffuse
flux of high energy neutrinos.

We select
events with high multiplicity of hit channels 
corresponding to bright cascades. The 
volume considered for generation of cascades is essentially
{\it below} the geometrical volume of {\it NT-96}.
A cut is applied
which accepts only time patterns corresponding to upward
traveling light signals (see below). 
Only the fewer atmospheric muons with large zenith
angles may escape detection and illuminate the array
exclusively via bright cascades below the detector. 
These events then have to be rejected by a stringent 
multiplicity cut.

Neutrinos produce showers and high energy muons through 
CC-interactions

\vspace{-0.4cm}
\begin{equation}
\nu_l(\bar{\nu_l}) + N \stackrel{CC}{\longrightarrow} l^-(l^+) + 
\mbox{hadrons},
\end{equation}
through NC-interactions

\vspace{-0.4cm}
\begin{equation}
\nu_l(\bar{\nu_l}) + N \stackrel{NC}{\longrightarrow} 
\nu_l(\bar{\nu_l}) + \mbox{hadrons},
\end{equation}
where $l=e$ or $\mu$, and through resonance production \cite{Glash,Ber1,Gandi}

\vspace{-0.4cm}
\begin{equation}
\bar{\nu_e} + e^- \rightarrow W^- \rightarrow \mbox{anything},
\end{equation}

\noindent
with the resonant neutrino energy  
$E_0=M^{2}_w/2m_e=6.3\cdot 10^6 \,$GeV 
and cross section $5.02\cdot 10^{-31}$cm$^2$.
\vspace{-0.4cm}
\begin{figure}
\centering
\mbox{\epsfig{file=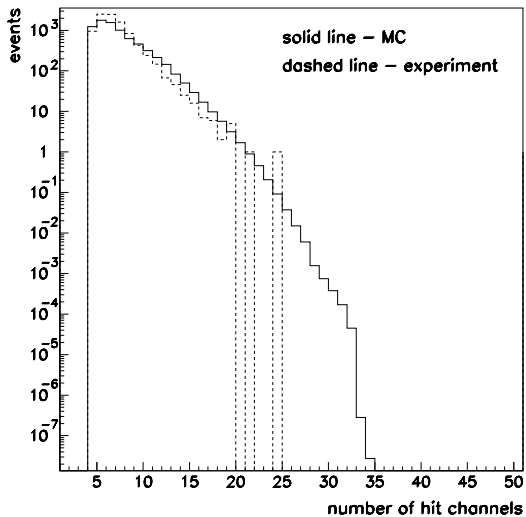,width=7.7cm}}
\caption { 
Hit channels multiplicity: solid histogram - showers produced by
atmospheric muons (MC), dashed histogram - experiment.
}
\end{figure}

\section{Detector}
The deep underwater Cherenkov detector {\it NT-96},
an intermediate stage of the Baikal Neutrino Telescope   
{\it NT-200}, has been operated since April 1996
till March 1997 \cite{APP2,JF}.
The detector comprises 96 optical modules (OM) at 4
vertical strings. 
The OMs are grouped in pairs along the strings. They contain 
37-cm diameter {\it QUASAR} PMTs 
\cite{Project,APP,OM2}. The two PMTs of a
pair are switched in coincidence in order to suppress light background
and PMT noise. A pair defines a {\it channel}. 
A {\it muon-trigger}
is formed by the requirement of \mbox{$\geq N$ {\it hits}}
(with {\it hit} referring to a channel) within \mbox{500 ns}.
$N$ is typically set to 
\mbox{3 or 4.} For  such  events, amplitude and time of all fired
channels are digitized and sent to shore. 

\vspace{-0.4cm}
\section{The data}
Within the first 70 days of effective data taking, $8.4 \cdot 10^7$ events
with the muon trigger $N_{hit} \ge 4$ have been selected. 
For this analysis we used events with $\ge$4 hits along at least one
of all hit strings. The time difference between any two channels
deployed on the same string was required to obey the condition:

\vspace{-0.4cm}
\begin{equation}
\mid(t_i-t_j)-z_{ij}/c\mid<a\cdot z_{ij} + 2\delta, \,\,\, (i<j),
\end{equation}
with $\delta=5$nsec and $a=1$ nsec/m.
The $t_i, \, t_j$ are the arrival times at channels $i,j$, and
$z_{ij}$ is their vertical distance. 
This condition has been used for almost vertically
up-going muons selection earlier \cite{APP2,JF}.

8608 events survive the selection criterion (4).
Fig.1 shows the hit multiplicity distribution
for these events (dashed histogram)
as well as the expected one for background showers
produced by atmospheric muons (solid histogram).
The experimental distribution is consistent with the theoretical
expectation within a factor 2. 
This difference can be explained by the uncertainty
of the atmospheric muon flux close to horizon at the detector 
\mbox{depth \cite{APP},} and by uncertainties in the dead-time of 
individual channels.

Since no events with $N_{hit}>24$ are found in our data we can derive
upper limits on the flux of high energy neutrinos which produce 
the events with multiplicity $N_{hit}>25$.

The energy dependences of the effective volumes for isotropic electron
and muon neutrinos 
are shown in Fig.2 (solid lines).
Also shown are the effective volumes folded with the neutrino
absorption probability in the Earth (dashed lines).

\section{The limits to the diffuse neutrino flux}
The shape of the neutrino spectrum was assumed to behave like 
$E^{-2}$ as typically expected for Fermi acceleration.
\begin{figure}
\centering
\mbox{\epsfig{file=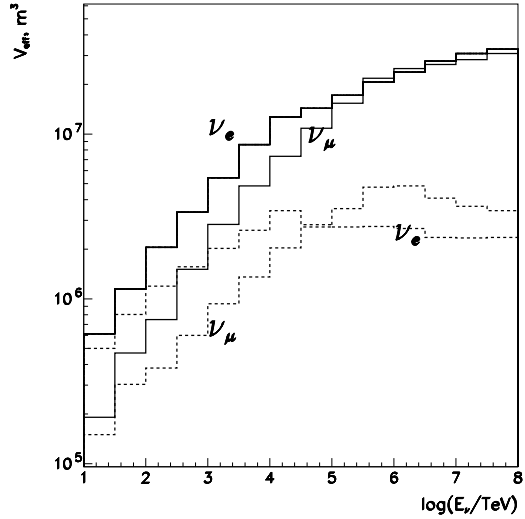,width=7.7cm,height=7.7cm}}
\caption { 
Effective volumes of {\it NT-96} for isotropic electron and muon neutrinos
(solid lines). 
The dashed lines represent the effective volumes folded with 
the neutrino 
absorption probability in the Earth.
}
\end{figure}
In this case, 90\% of expected events would be produced by  neutrinos
from the energy
range $10^4 \div 10^7$GeV with the center of gravity around $2 \cdot 10^5$GeV.
Comparing the calculated rates with the upper limit to the 
actual number of 
events, 2.3 for 90\% CL, and
assuming the flavor ratios $\Phi_{\nu_{\mu}}=\Phi_{\bar{\nu_{\mu}}}=
\Phi_{\nu_e}$ due to photo-meson production of $\pi^+$ followed by
the decay $\pi^+ \rightarrow \mu^+ + \nu_{\mu} \rightarrow 
e^+ + \nu_e + \bar{\nu_{\mu}} +\nu_{\mu}$  
for extraterrestrial sources \cite{BAHC,P98},
we obtain the following upper
limit to the diffuse neutrino flux:
\begin{equation}
\frac{d\Phi_{\nu}}{dE}E^2<1.4\cdot10^{-5} 
\mbox{cm}^{-2}\mbox{s}^{-1}\mbox{sr}^{-1}\mbox{GeV}.
\end{equation}

New theoretical upper bounds to the intensity of high-energy
neutrinos from extraterrestrial sources have been presented 
recently \cite{BAHC,P98}.
These upper bounds as well as our limit and limits obtained
by DUMAND \cite{DUMAND}, AMANDA \cite{AMANDA}, EAS-TOP \cite{EAS}
and FREJUS \cite{FREJUS} experiments are shown in Fig.3.
Also, the atmospheric neutrino \mbox{fluxes \cite{LIP}} 
from horizontal  and vertical directions (upper and lower curves,
respectively) 
are presented.

\vspace{-0.4cm}
\begin{figure}
\centering
\mbox{\epsfig{file=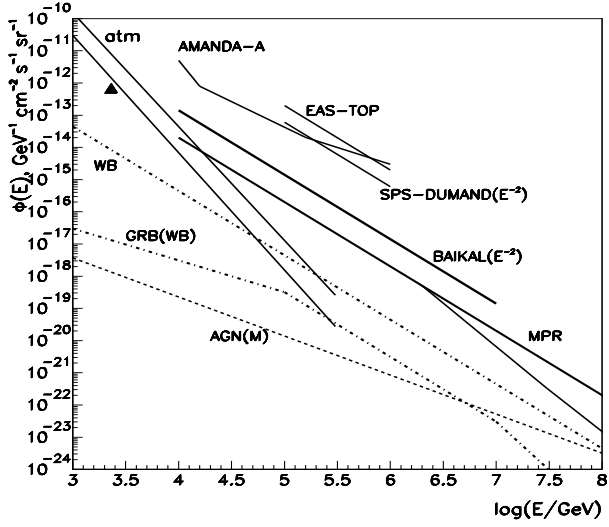,width=8.8cm,height=7.5cm}}
\caption { 
Upper limits to the differential flux of high 
energy neutrinos obtained by 
different experiments as well as upper bounds for  
neutrino fluxes from a number of different models. 
Dot-dash curves
labeled WB and GRB(WB) - upper bound and neutrino intensity
from GRB estimated by Waxman and Bahcall (1997,1999);
dashed curve labeled AGN(M) - neutrino intensity from AGN (Mannheim 
model A, 1996);
solid curves labeled MPR - upper bounds for $\nu_{\mu}+\bar{\nu_{\mu}}$
in Mannheim et al. (1998) for pion photo-production neutrino sources
with different optical depth $\tau$ (adapted from ref.17).  
The triangle denotes the limit obtained by the Frejus
for an energy of 2.6 TeV :  $7 \cdot 10^{-13} \mbox{cm}^{-2}
\mbox{s}^{-1} \mbox{sr}^{-1} \mbox{GeV}^{-1}$.
}
\end{figure}

\newpage
For resonant process (3) our 90\% CL limit is:
\begin{equation}
\frac{d\Phi_{\bar{\nu}}}{dE_{\bar{\nu}}} \leq 3.6 \times 
10^{-18} 
\mbox{cm}^{-2}\mbox{s}^{-1}\mbox{sr}^{-1}\mbox{GeV}^{-1}.
\end{equation}
This limit lies between  limits obtained by DUMAND ($1.1 \times 10^{-18}$
cm$^{-2}$s$^{-1}$sr$^{-1}$GeV$^{-1}$) and EAS-TOP 
($7.6 \times 10^{-18}$cm$^{-2}$s$^{-1}$sr$^{-1}$GeV$^{-1}$). 

The new limits (5) and (6) have been obtained with underwater
detector {\it NT-96}.
Analysis of 3 years
data taking with {\it NT-200} \cite{APP2,JF} would allow us to lower 
this limit by another order of magnitude.

\bigskip

{\it This work was supported by the Russian Ministry of Research,the German 
Ministry of Education and Research and the Russian Fund of Fundamental 
Research \mbox{( grants }} \mbox{\sf 99-02-18373a}, \mbox{\sf 97-02-17935}, 
\mbox{\sf 99-02-31006}, \mbox{\sf 97-02-96589} {\it and} \mbox{\sf 97-05-96466}).

\end{document}